\documentclass[nature,twocolumn,superscriptaddress,floatfix,showpacs,nofootinbib]{revtex4-1}
\usepackage{graphics,amssymb,amsmath,epsfig,color}

\newcommand{\be}{\begin{equation}} \newcommand{\ee}{\end{equation}}
\newcommand{\bea}{\begin{eqnarray}} \newcommand{\eea}{\end{eqnarray}}

\begin{document}

\title{Discontinuous Percolation Transitions in Epidemic Processes, Surface Depinning in Random Media and 
Hamiltonian Random Graphs}
\author{Golnoosh Bizhani} \affiliation{Complexity Science Group, University of Calgary, Calgary T2N 1N4, Canada}
\author{Maya Paczuski} \affiliation{Complexity Science Group, University of Calgary, Calgary T2N 1N4, Canada}
\author{Peter Grassberger} \affiliation{Complexity Science Group, University of Calgary, Calgary T2N 1N4, Canada}
\date{\today}

\begin{abstract}

Discontinuous percolation transitions and the associated tricritical 
points are manifest in a wide range of  both equilibrium and non-equilibrium 
cooperative phenomena.  To demonstrate this,
we present and relate the continuous and first order behaviors in two
different classes of models: The first are 
generalized epidemic processes (GEP) that describe in their spatially embedded version
-- either on or off a regular lattice -- compact or fractal cluster growth
in random media at zero temperature. A random graph version of these processes is mapped
onto a  model  previously proposed for complex social contagion. We compute detailed
phase diagrams and compare our numerical results at the tricritical point
in $d=3$ with field theory predictions of Janssen {\it et al.} 
[Phys. Rev. E {\bf 70}, 026114 (2004)]. The second class consists of exponential
(``Hamiltonian", i.e. formally equilibrium) random graph models and includes the 
Strauss and the 2-star model, where `chemical potentials' control the densities of 
links, triangles or 2-stars. When the chemical potentials in either graph model are 
${\cal O}(\log N)$, the percolation transition can coincide with a
first order phase transition in the density of links, making the former 
also discontinuous. Hysteresis loops can then
be of mixed order, with second order behavior for decreasing link 
fugacity, and a jump (first order) when it increases.

\end{abstract}

\maketitle

\section{Introduction}

Percolation describes the sudden appearance of system-wide connectivity arising from 
microscopic processes. It is a classic example~\cite{Stau-1994} of a continuous 
(``second order'') phase transition. Interest in systems where the percolation 
transition is discontinuous (``first order") was sparked recently by claims for this 
in {\it Achlioptas processes}~\cite{Achli-2009}. Although these transitions were 
later shown to be continuous~\cite{Costa-2010,Riordan,Grassberger,Park-2011} -- 
albeit with unusual finite size scaling behavior~\cite{Grassberger}, the fact that 
percolation transitions could be discontinuous was claimed to be novel or even 
revolutionary~\cite{vespignani2010fragility,Friedman-2009,Ziff-2009,Ziff-2010,Radicchi-2009,Radicchi-2010,Souza-2010,Rozenf-2010,Manna-2009,Moreira-2010,Araujo-2010,Chen-2010,Cho-2009,Cho-2009a,Cho-2010,Tian-2010,Nagler-2011,Hooy-2011,Costa-2010,Riordan}.
After this, discontinuous percolation transitions were observed in 
interdependent
networks~\cite{Buldy-2010,Parshani-2010,Son-2011}, in models inspired 
by~\cite{Achli-2009}
but not using the Achlioptas 
trick~\cite{Manna-2009,Moreira-2010,Araujo-2010,Chen-2010},
and in a hierarchical lattice~\cite{Boettcher-2011}.

One purpose of this work is to point out that discontinuous percolation transitions 
are not surprising and, indeed, are common to a variety of (e.g. social or physical) 
cooperative phenomena. The existence of such transitions, together with an associated 
tricritical point, was proposed 25 years ago~\cite{Ohtsuki-1987b} in the context of 
directed percolation. This was verified in the seminal field theoretic work of 
Janssen {\it et al.}~\cite{Janssen-2004} who introduced the {\it generalized epidemic 
process} (GEP) \footnote{Janssen {\it et al.} called this the 'generalized general
epidemic process' (GGEP), in order to distinguish it from the 'general epidemic process'
[D. Mollison, J. Royal Statist. Soc. {\bf B} 39, 283 (1977)]. The latter is now 
usually called an SIR (susceptible-infected-removed) epidemic, while a 'simple epidemic' 
in the notation of Mollison is now called SIS. We feel thus free to use the 
simpler acronym GEP for the generalized process.}. In this scenario, the 
continuous transition is just ordinary percolation (OP), whilst the discontinuous one 
is the depinning  transition of driven surfaces in random media at zero
temperature~\cite{Cieplak-1988,Martys-1991,Martys-1991b, Ji-1992,Nolle-1993,Hecht-2004,Drossel}.

Although the latter is continuous from the point of view of surface properties, it is
discontinuous so far as the percolation order parameter is concerned~\footnote{Note 
that the co-appearance of first order bulk transitions with
continuous surface transitions was discussed already in~\cite{Lipowsky}.}. 
Indeed, the density of `wetted' sites in the presence of a driven interface 
jumps discontinuously from zero to a finite value at depinning.

A closely related line of papers finding discontinuous percolation 
transitions started independently
in a social science context~\cite{Watts-2002,Dodds-2004,Dodds-2005} and 
addresses complex contagion or
epidemics in random networks. It turns out that the model 
of~\cite{Dodds-2004} is basically
the random graph version of the GEP, as we explain in detail below. 
Our unified formulation based on GEP, that includes both social contagion and
interface depinning, simplifies the description of both and isolates relevant 
variables that can affect the actual outcome in terms of potentially 
measurable observables.

We present  numerical simulation results for the GEP, including the
time dependence of the number of growth sites in three dimensions. At 
the first order (= depinning) transition line, activity decays as a stretched 
exponential in time whilst it behaves as a power-law both at the OP 
transition and at the tricritical point which separates rough from
fractal growth. We give (tri-)critical exponents and compare them to 
theoretical predictions~\cite{Janssen-2004}.

The main property that leads to first order transitions in the models we 
consider is cooperativity (or `synergy') in establishing links. This 
cooperativity can be implemented technically in different ways.
We do this  via stochastic
dynamics as in~\cite{Janssen-2004,Watts-2002,Dodds-2004,Dodds-2005} for 
the GEP, and also via Gibbs-Boltzmann {\it equilibrium} distributions in
Hamiltonian (or ``exponential") ensembles, which have been used 
extensively to model social networks.

Indeed, we also find discontinuous percolation transitions
in two exponential random graph models :  the 
Strauss~\cite{Strauss-1986,Park-2005}
and the 2-star model~\cite{Park-2004a}. They are both formulated in 
terms of a partition function and are generalizations of the standard 
Erd\"os-Renyi (ER) random graph~\cite{Boll-1985}.  The Hamiltonians are 
bilinear with a control parameter ($\theta$) conjugate to the number of 
links and another control parameter conjugate to either the number of 
triangles or the number of ``2-stars.'' When all control parameters
are ${\cal O}({\log N})$ (where $N$ is the number of nodes in the 
graph), the percolation transition can be either continuous or 
discontinuous, with  hysteresis loops typical of first-order 
transitions. But for certain parameter regimes unusual hysteresis 
loops occur, where the percolation order parameter exhibits second-order 
(singular but continuous) behavior for decreasing $\theta$, but jumps 
discontinuously for increasing $\theta$.  Such ``mixed-order"
hysteresis loops have not to our knowledge been seen before.

It is  well known that the observation of continuity of a phase transition depends 
not only on the choice of the order parameter, but also on the choice of the control parameter. 
Take e.g. the standard example of a liquid-gas transition. If the temperature of water is increased 
at constant pressure, then the density and the free energy jump discontinuously at the boiling 
temperature. If, however, the volume is kept fixed, no such jump is observed. Instead, as 
temperature is increased, a larger and larger fraction of the sample turns into vapor, but this 
happens in a completely continuous way. The standard assumption in thermostatics is that the 
order parameter is a density or inverse density (e.g. specific volume), and the control 
parameter is its conjugate (e.g. pressure).  But in percolation, the standard choice of order 
parameter is the fraction $S_{\rm max}/N$ of sites belonging to the giant cluster, while the 
control parameter is usually also a density -- the density of occupied sites  (bonds) in site 
(bond) percolation. Although this choice is legitimate, it can obscure the notion of first 
{\it vs.} second order transitions, since other choices more in line with thermostatics can lead 
to different conclusions. This might explain why previous works on first order percolation 
transitions were not recognized as such in the recent literature.

\section{The Generalized Epidemic Model: Complex Contagion Treated as a Stochastic Process}

Although the epidemic model of Ref.~\cite{Janssen-2004} is formulated as a continuum field 
theory, the situation becomes more clear on a lattice. Consider a process where the probability
of a given site becoming infected (or invaded) by one of its neighbors depends on the number 
of previous attempts by other neighbors. Once a site is infected, it tries once to infect every one of 
its not yet infected neighbors.  Denote by $p_k$ the probability that an infection
succeeds, if the attacked site has already fended off $k$ previous attacks. If every attack
increases the strength of the defender, $p_k$ decreases with $k$, otherwise (if it weakens
it), $p_k$ increases. Site percolation is described by $p_0 >0$ and $p_k=0$ for $k\geq 1$: 
If the first attack does not succeed, all later attempts are futile. Bond percolation is described by 
$p_k=p$ for all $k\geq 0$. Ordinary (second order) percolation is observed whenever $p_k$ decreases 
with $k$ (see also ~\cite{Perez-2011}), but the transition switches to first order when $p_k$ 
increases sufficiently fast. In that case, infected clusters fill in most 
holes and bays, while protrusions are avoided -- thereby making the clusters compact with 
rough but non-fractal surfaces. (The same effect is caused by high surface tension 
compared to disorder  at the cluster-void interface in random media). Detailed predictions 
for the tricritical behavior in terms of an $\epsilon  = 5 -d$ expansion were 
given in~\cite{Janssen-2004}. 

\subsection{The GEP on Random Graphs}

Although Dodds {\it et al.}~\cite{Dodds-2004,Dodds-2005} assume a somewhat more complex mechanism of 
infection, their basic model can be mapped onto a sparse random graph model where each node 
with $n$ neighbors in the the giant cluster is itself in the giant cluster with probability $q_n$. This 
is precisely the mean field (random graph) version of the above model, if $q_1=p_0$ and $q_{n+1}=q_n
+(1-q_n)p_n$ \footnote{Strictly spoken this is only true for static (i.e., percolation in the narrow 
sense) aspects at threshold. Dynamics and the behavior above the threshold are different in both models, 
since the memory about previous contacts with infected neighbors is short lived in \cite{Dodds-2004,Dodds-2005}, 
while it is long lived (previous attacks are never forgotten) in the model of \cite{Janssen-2004} and 
in the simulations reported in the next subsection.}. Due to the absence of short loops in this case, 
the condition for tricriticality (transition between classes I and II in~\cite{Dodds-2004,Dodds-2005}) simplifies to
\be
q_2 = 2 q_1,    \label{qq}
\ee
with no restriction on any $q_n$ with $n\geq 3$. Since the derivation of this 
in~\cite{Dodds-2004,Dodds-2005}
is somewhat involved and obscures the relationship to the GEP as defined in \cite{Janssen-2004},
a simple proof of Eq.(\ref{qq}) is given in the supplementary material.

\subsection{The GEP on Regular Lattices: Tricritical Behavior and Rough Pinned Surfaces}

In the present work we studied in detail the case where $p_k\equiv p$ is the same for all $k>0$, 
while $p_0$ is different. Phase diagrams for simple (hyper-) cubic lattices and for random regular
graphs are shown in Fig.~1. To the left of the curves, no infinite clusters exist, while such clusters
do exist to their right. Since percolation thresholds on lattices scale as $p_c \sim 1/(2d-1)$ for 
large $d$, we used $(z-1)q_1$ and $(z-1)q_2$ as coordinates in Fig.~1, where $z$ is the coordination 
number. All curves start at site percolation ($p=0,\; q_1=q_2=p_0$), since we do not 
consider here antagonistics effects (i.e., two attacks together cannot have less success than a 
single one). Tricritical points are marked by circles. There is no tricricital transition in $d=2$
\cite{Drossel}, i.e. isotropic rough 1-$d$ surfaces are always fractal. For large dimensions, the 
lattice results converge to those for regular random graphs with degree $z = 2d$, as short loops
become less and less important with increasing $d$. Preliminary results indicate that surface properties
in the first-order regimes (i.e. below the tricritical points, but for $q_1>0$) are {\it not} in 
the same universality class of surfaces without overhangs, which is the accepted theory for pinned rough surfaces~\cite{Leschhorn,Doussal,Rosso}, while
the situation is less clear in the case $q_1=0$, i.e. if at least two infected neighbors are needed for a site
to become infected (in the latter case, epidemics can neither spread from single sites nor from $(1,0,0\ldots)$
surfaces, but they can spread from $(1,1,1\ldots)$ surfaces).

\begin{figure}
\psfig{file=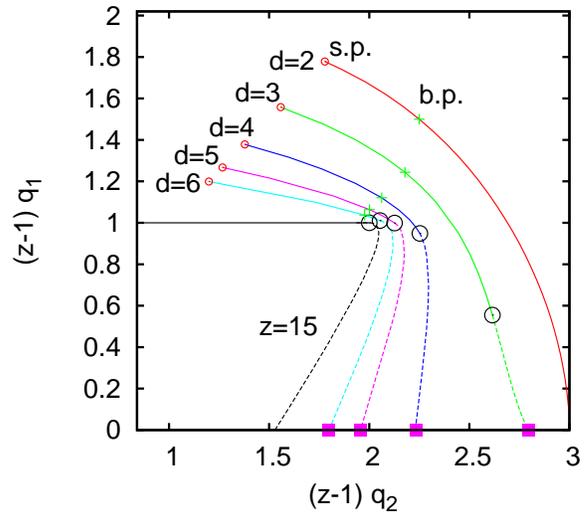,width=7.5cm, angle=270}
\caption{(Color online) Phase diagrams for the generalized epidemic process where infection of a site succeeds with
probability $p_0$ in the first encounter with an infected neighbor, while it succeeds with chance $p$ in all
later encounters. Following ~\cite{Dodds-2004}, we use instead of $p_k$ the probabilities $q_1=p_0$ and $q_2 
=p_0+(1-p_0)p$ that that an infection has occurred after two encounters. Since percolation thresholds
on regular graphs are roughly $\propto 1/(z-1)$ for large degree $z$, we use actually $(z-1)q_n$ for the 
two axes. The curves labeled by $d=2$ to $d=6$ correspond to $d-$dimensional hypercubic lattices, where 
$z=2d$, while the curve labeled ``z=15" is for random (i.e. locally loopless) networks. All curves start
at the site percolation point $q_2=q_1$ (small red circles). The bond percolation points (green crosses)
are at $q_2=(2-q_1)q_1$. For large $d$ they approach the tricritical points (big black circles) that 
converge as $((k-1)q_1,(k-1)q_2) \to (1,2)$ for large $k$. The percolation transitions are first order 
below the tricritical points, and second order above. For $d=2$ there is no tricritical point, i.e. 
clusters and their surfaces are always fractal. In the first order regime (dashed curves), surfaces develop strong 
overhangs and seem {\it not} to be described by models where these overhangs are neglected, except 
possibly at $q_1=0$ (magenta squares), which seem to represent a different fixed point for $d=3$.}
   \label{GEP}
\end{figure}

\begin{figure}
\psfig{file=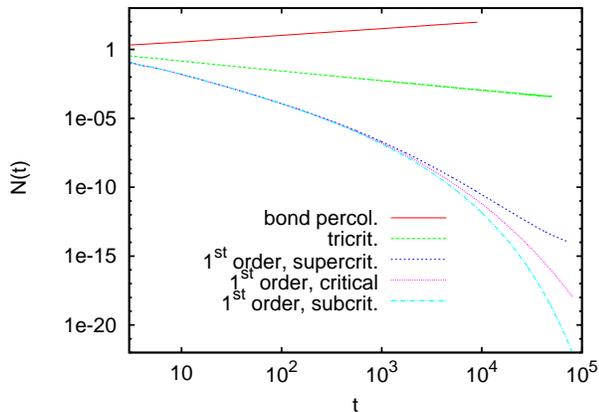,width=5.8cm, angle=270}
\caption{(Color online) Time dependence of the activity $N(t)$ at five different pairs $(p_0,p)$ 
where $p_k=p$ for all $k\geq 1$, for the generalized epidemic process on the simple 3-d cubic lattice
starting from a single infected site.
The uppermost (red) curve is for $p_0=p=0.2488\ldots$ which is critical bond percolation. The middle
(green) curve is for the tricritical point $(p_0,p)=(0.111,0.464)$ (i.e., $(q_1,q_2) = (0.111,0.523)$). 
Within the resolution of the 
figure, both these lines are straight. The lowest three curves are near the first order (depinning)
transition $(p_0,p)=(0.07,0.792318+\Delta)$, where $\Delta = 0$ (magenta), $+0.00022$ (dark blue) 
and $-0.00022$ (light blue). Statistical error bars are all smaller than the 
line widths.
   }
   \label{tricritical}
\end{figure}

For the simple cubic lattice, simulations show that the tricritical point for this model is at at $p_0=0.111(2)$ 
and $p_k= 0.464(8)$ for $k>0$. Results of such simulations for epidemics starting from a single 
infected site are shown in Fig.~\ref{tricritical}, where $N(t)$ is the number of sites newly 
infected at time $t$. For OP, $N(t)$ increases as a power law $t^{\eta}$ with $\eta \approx 0.35$ 
\cite{Stau-1994}, it decreases at the tricritical point as $N(t) \sim t^{\eta_s}$ with 
$\eta_s = -0.70(1)$. This is in stark contrast to the prediction $\eta_s \approx 0.05$ of 
\cite{Janssen-2004}.
The tricritical point and OP are the only cases where $N(t)$ shows a power law. For critical 
percolation with $p_0>0.111$ the behavior crosses over to the OP scaling, while for $p_0<0.111$
the data are compatible with a stretched exponential at the transition line (see lowest three curves in 
Fig.~\ref{tricritical}). Analogous plots for the probability $P(t)$ that the epidemic survives
at least $t$ time steps and for its average squared radius $R^2(t)$ are given in the supplementary
material. They give $\delta_s = 1.49(2)$ and $z_s = 1.205(4)$, where $\delta_s$ and $z_s$ are 
defined via $P(t)\sim t^{-\delta_s}$ and $R^2(t) \sim t^{z_s}$. The predictions of \cite{Janssen-2004}
are $\delta_s \approx 0.87$ and $z_s \approx 1.06$. Again the agreement is far from perfect, although
the changes from the OP critical exponents are in the right directions. More details are given 
in the supplementary material and in \cite{grass-unpub}. For the completely analogous case of 
directed percolation (SIS epidemics), see \cite{Janssen-2005,Lubeck-2006,Grass-2006}.

\section{Cooperative Percolation in Hamiltonian Random Graph Models}

The above discussion suggests that cooperativity in finite temperature equilibrium systems 
may also lead to discontinuous percolation transitions. Indeed, the mean field percolation transition 
corresponds to the emergence of the giant component in ER random graphs, where links appear 
{\em independently} with probability $p$~ \cite{Boll-1985}. The latter is the simplest  
``exponential model"~\cite{Holland-1981,Park-2004,Robins-2007}. In this approach one considers  
graphs $G$ with  $N$  nodes, were the probability for a given graph 
is defined by the Boltzmann-Gibbs  equilibrium formula
\be
   P(G;\theta_1,\theta_2,\ldots) = {1\over Z} e^{-H(G;\theta_1,\theta_2,\ldots)}
\ee
with $Z = \sum_G e^{-H(G;\theta_1,\theta_2,\ldots)}$. Here $H$ is 
the Hamiltonian,  $\{\theta_1,\theta_2,\ldots\}$ represents a set of control parameters, and we have set $\beta\equiv 1/kT=1$.
More precisely, we assume that $H$ is a sum of bilinear terms,
\be
   H(G;\theta_1,\theta_2,\ldots) = \sum_\alpha \theta_\alpha A_\alpha(G) \;,
\ee
where each $A_\alpha$ is an observable (``statistic") of the graph, and $\theta_\alpha$ 
is the associated chemical potential.  Typically, each $A_\alpha$ represents the total number of  small
subgraphs (links, triangles, $p$-stars, 4-cliques, ...) in the graph.

Models of this type have been popular in mathematical sociology \cite{Strauss-1986,
Robins-2007}, although they tend to be unrealistic.  In many cases, such models reduce to equivalent ER graphs without clustering and are trivial, apart 
from the usual, non-trivial dependence of the observables  $A_\alpha$ on the control parameters $\theta_\alpha$ ~\cite{Chatter-2011}.

 The Hamiltonian for the ER model is: \be
   H_{\rm ER}(G;\theta) = \theta L(G)\;,
\ee
where $L(G)$ is the number of links in $G$ and $\theta = \ln [(1-p)/p]$. It exhibits
 a percolation transition at $p=1/N$ (when $N\to\infty$)~\cite{Boll-1985}, thus the critical value of $\theta$ is 
\be
   \theta_c = \ln N.
\ee
In the following, we study the 2-star model~\cite{Park-2004a} with
\be
   H_{\rm 2-star}(G;\theta,J) = \theta L(G) -\frac{J}{N} n_2(G)\;.  \label{star}
\ee
Here $n_2(G)$ is the total number of ``2-stars", i.e. of pairs of links attached to the 
same node.  We also consider the Strauss model~\cite{Strauss-1986,Park-2005} with
\be
   H_{\rm Strauss}(G;\theta,B) = \theta L(G) -\frac{B}{N} n_\Delta(G)\;,  \label{strauss}
\ee
where $n_\Delta(G)$ is the total number of distinct triangles, i.e. of loops of length 3. In 
terms of the degree sequence $\{k_i, i = 1\ldots N\}$, 
\be
   L(G) = \frac{1}{2}\sum_{i\in G}k_i, \quad {\rm and} \quad 
             n_2(G) = \frac{1}{2}\sum_{i\in G}k_i(k_i-1),
\ee
while $n_\Delta(G)$ depends also on degree correlations.

For a typical (non-sparse) graph $L$ increases quadratically with $N$, while both $n_2$
and $n_\Delta$  $\sim N^3$. This is why $J/N$ and $B/N$ are used as 
control parameters in Eqs.~(\ref{star},\ref{strauss}) instead of $J$ and $B$.  The 2-star model has, for any $\theta > 2$, a first
order transition in the density of links at $J^*(\theta)$ and  strong hysteresis.  The results of~\cite{Park-2004a},
together with a standard Maxwell construction  show that \be
   J^*(\theta) = \theta \quad .
\ee
The line of first order transitions terminates at the critical point $\theta^*_c=J^*_c=2$. 
(We neglect here all terms that are ${\cal O}(1/N)$ relative to the leading ones). Similarly, 
for the Strauss model a first order transition 
in the link density $p=\langle L \rangle/N$ occurs for any $\theta \gtrsim 0.81$~\cite{Park-2005}. 
This time it is more complicated to obtain the exact transition line $B^*(\theta)$, 
but one can show that (see Supplementary Material)
\be
   B^*(\theta) \approx 3 \theta \quad {\rm for} \quad \theta \gg 1,
\ee
with a critical point  at $p^*_c = 2/3, B^*_c = 27/8 = 3.375$, and $\theta^*_c =3/2-\ln(2) \approx 0.807$.

For both models, a giant  component exists in both  the high  and 
low link density phases, whenever $\theta = {\cal O}(1)$. Thus the density transition
happens  when they are already percolating, as long as $\theta$ is finite. In 
order to reach a percolation transition one has to take $\theta \sim \ln N$ to get a sparse graph. In this 
regime the above estimates for the density transitions are still valid. Moreover, 
when $B<B^*$ or $J<J^*$, respectively, the second terms in the Hamiltonians (Eqs.~(\ref{star},\ref{strauss})) have no influence on the percolation transition, for $N\to\infty$. This is 
illustrated for the 2-star model in Fig.~2, where we show numerical results averaged
over 50 hysteresis loops for a network with $N=2000$ and $J = 3.0$. The 
density transition (monitored via the average degree) indeed appears to be  first order 
with strong hysteresis, while the percolation transition (monitored via $\langle
S_{\rm max}\rangle/N$) shows  no hysteresis and is {\it exactly} the same as for 
ordinary ER networks.

\begin{figure}
\psfig{file=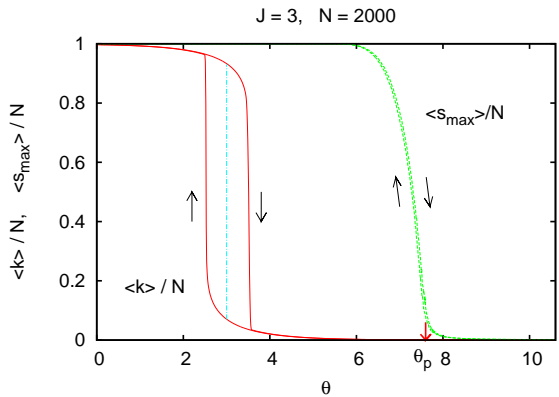,width=5.5cm, angle=270}
\caption{(Color online) Hysteresis loops for the 2-star model when holding the chemical 
   potential $J$ fixed at a value of ${\cal O}(1)$ and sweeping the control
   parameter $\theta$ for links. The loop on the left (red) is for the normalized 
   average degree, while the one on the right (green) is for the percolation order 
   parameter $\langle S_{\rm max}\rangle/N$. The vertical dashed (blue) line is the Maxwell 
   prediction for the true density transition. The curve for $\langle S_{\rm max}\rangle/N$
   shows practically no hysteresis and agrees within error with the one for OP. 
   The percolation threshold $\theta_c = \ln N$ is indicated on the x-axis. 
   The rounding of the green curve near $\theta_c$ is a finite size effect ($N=2000$).
   }
   \label{fig1}
\end{figure}

\begin{figure}
\psfig{file=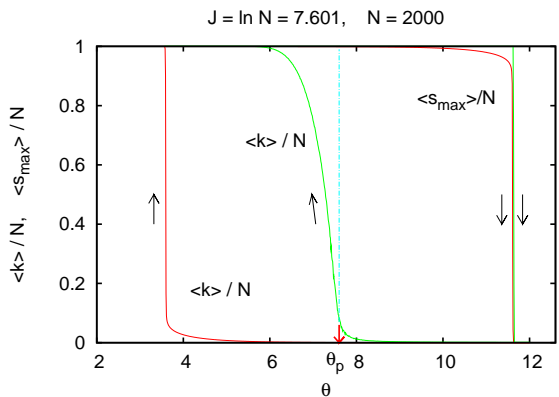,width=5.5cm, angle=270}
\caption{(Color online) Analogous to Fig.~3, but for $J = \theta_c = \ln N = 
   7.601$. This time the percolation threshold coincides with the true (Maxwell) 
   density transition. As a consequence, also the curve for $\langle S_{\rm max}\rangle/N$
   shows hysteresis, with a continuous lower branch and a discontinuous upper branch.
   }
   \label{fig2}
\end{figure}

When $J > \ln N$ (or $B > \ln N$, respectively) this scenario breaks down because
the true equilibrium state at the ER percolation threshold, $\theta_p$, is a dense graph that consists of a 
single giant component. Although the equilibrium network percolates at $\theta_p$,
a hysteresis loop starting at $\theta > J$ 
begins with a sparse non-percolating graph, and due to metastability the effect of 
$J$ (or $B$) is not seen  until one passes the ER percolation threshold -- provided
that it remains in the metastable region. 
This scenario is illustrated in Fig.~2 for $J = \theta_p = \ln N$. Now the 
hysteresis loop for $\langle k\rangle/N$ is quite wide. The hysteresis loop for 
$\langle S_{\rm max}\rangle/N$ shows the ordinary ER percolation shape on its lower branch,
while it follows the discontinuous behavior of $\langle k\rangle/N$ on the upper
(return) branch. To our knowledge, such a mixed-order hysteresis loop has not been 
observed before.

\section{Conclusions}

As we already mentioned  and as was pointed out repeatedly 
before~\cite{Park-2004a,Park-2005,Chatter-2011}, 
the 2-star and Strauss models are not realistic for real world applications.
Accordingly, our demonstration that they exhibit both  second and first order 
percolation transitions should be considered only as a proof that this phenomenon exists in 
equilibrium. More interesting examples can be easily suggested.
A class of models that come into mind are random graphs (i.e. mean field type) 
where not only the number of nodes but also the number of links is set by
hard constraints (``microcanonical models"). A model with two control parameters
($B$ and $J$) studied in~\cite{Golnoosh-2012} exhibits
zoo of metastable states, but it seems that most of these states are still too 
extreme to be physical. A further step in this direction could be to fix not only
the average degree, but to fix the entire degree distribution, either by soft 
\cite{Park-2011a} or by hard \cite{Foster-2010} constraints. 

A more realistic class of Hamiltonian models with first order phase 
transitions could be spatially embedded (e.g. finite dimensional lattice) systems.
Such models have not been studied much in the social science literature, although
it is  well-known that, for instance, spatial structure is essential to maintain diversity
in ecosystems. We believe that such models might provide a suitable
mixture of structure and randomness to reveal important features of real, complex networks, 
including presumably percolation transitions of both continuous and discontinuous 
type.

In summary, we have shown that various percolation models can be naturally
generalized such that they switch from ordinary, continuous behavior at the
transition point to  discontinuous (``first order") behavior, as some parameter is
varied. This parameter usually is a measure of cooperativity in linking or ``infecting" sites 
(the probability for sites to get linked is increased by other links already present), 
such that the percolation transition is more abrupt when cooperativity is high. We 
present a unified treatment including examples that range from social dynamics to 
condensed matter physics, and we also show that analogous phenomena occur both in stochastic 
dynamics out of equilibrium as well as in a Gibbs-Boltzmann equilibrium framework.  
We also simplify the  dynamical description of the social contagion process introduced by 
Dodds and Watts \cite{Dodds-2004}, clarifying thereby its relation to percolation and to the 
generalized epidemic process defined in \cite{Janssen-2004}.
 
In condensed matter physics, the first order percolation transition is just the depinning
transition of driven interfaces in disordered media at zero temperature. Treating it also in our unified 
framework not only allows us to study in detail the tricritical point
(where we found for $d=3$ striking disagreement with theoretical predictions), but
also to numerically investigate more efficiently a model for pinned surfaces in which overhangs 
are fully included and hence the rotational symmetry of the growth process is not explicitly 
broken at scales much less than the system size.
In this latter context, the most important (but so far only preliminary) result we find is that
overhangs are indeed crucial for such surfaces, and that all existing theories for
critically pinned rough surfaces (which neglect overhangs and are based on a single-valued 
"height function") might be obsolete, not being relevant to the most interesting physical 
case of isotropic media.

%
%
\bibliographystyle{apsrev4-1}

\bibliography{mm}

\end{document}


\title{Supplementary Material, ``Discontinuous Percolation Transitions in Epidemic Processes, Surface Depinning in Random Media and Hamiltonian Random Graphs"}
\author{Golnoosh Bizhani} \affiliation{Complexity Science Group, University of Calgary, Calgary T2N 1N4, Canada}
\author{Maya Paczuski} \affiliation{Complexity Science Group, University of Calgary, Calgary T2N 1N4, Canada}
\author{Peter Grassberger} \affiliation{Complexity Science Group, University of Calgary, Calgary T2N 1N4, Canada}
\date{\today}

\maketitle

\section{Details on the Simulation of Tricritical and First Order Percolation on 3-d Lattices}

All simulations were done on simple cubic lattices, with synchronous (discrete time) updates. We followed the 
spreading of epidemics that started either with point seeds or with the seed consisting of an entire infected 
plane. In the former we used lattices of size up to $2048^3$ and checked that clusters never reached
the boundary. For small values of $p_0$, when growth from a point seed has a very high chance to die out, we 
used PERM \cite{PERM} to grow clusters even if their probability was as low as $10^{-300}$. For simulations 
initiated from an entire infected plane (results of which are not shown here but are used, in addition, to better
estimate (tri-)critical points) we used lattices of sizes up to $4096^2\times 2048$ with helical lateral 
boundary conditions. In that case we also implemented multi-spin coding in order to use only 2 bits to store 
the status of any site. We also recycled memory in order to grow epidemics that spread far from the initial infected 
boundary, by overwriting older parts of the cluster that were no longer growing. This enlarges the effective lattice 
size to $4096^2\times L_z$ with $L_z \gg 2048$. The precise thickness that we must not overwrite
depended of course on the actual roughness of the growing surface. It was always
checked that the cluster could grow without improperly interfering with some of its older parts.

\begin{figure}
\psfig{file=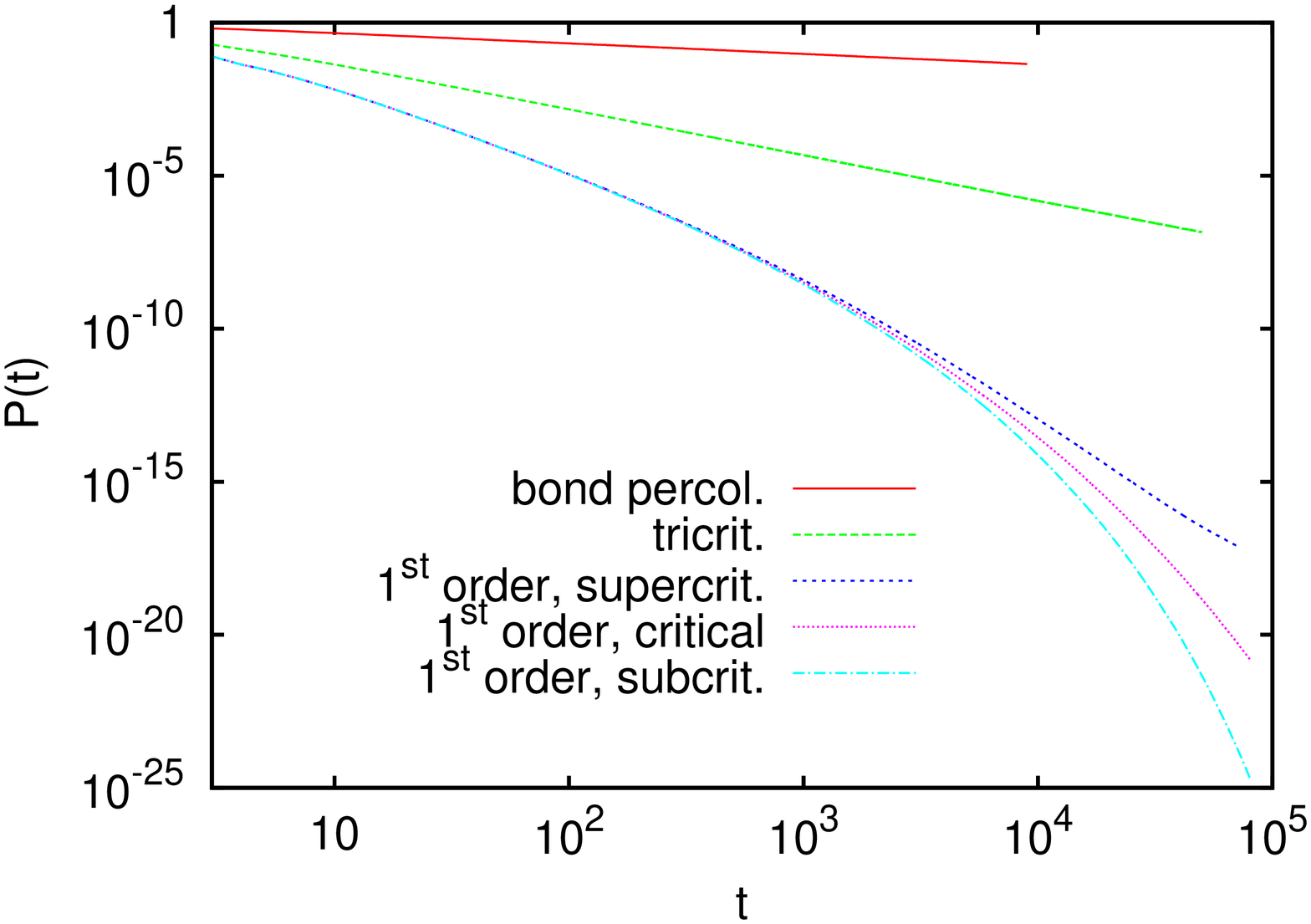,width=5.8cm, angle=270}
\caption{(Color online) Log-log plot analogous to Fig.~1 of the main text, but for $P(t)$ 
   instead of $N(t)$. Here $P(t)$ is the probability that an epidemic started which with a single 
   infected site is still growing after $t$ time steps. The decay follows a power law 
   both at OP and at the tricritical point (with exponent $-1.49(2)$ for the latter), while it 
   seems to follow a stretched exponential 
   at the first order (i.e., rough surface depinning) transition point. Again the central one
   of the three lowest curves (for $p_0=0.07$) corresponds to depinning.}
   \label{fig.S1}
   \vskip -.5 cm
\end{figure}

\begin{figure}
\psfig{file=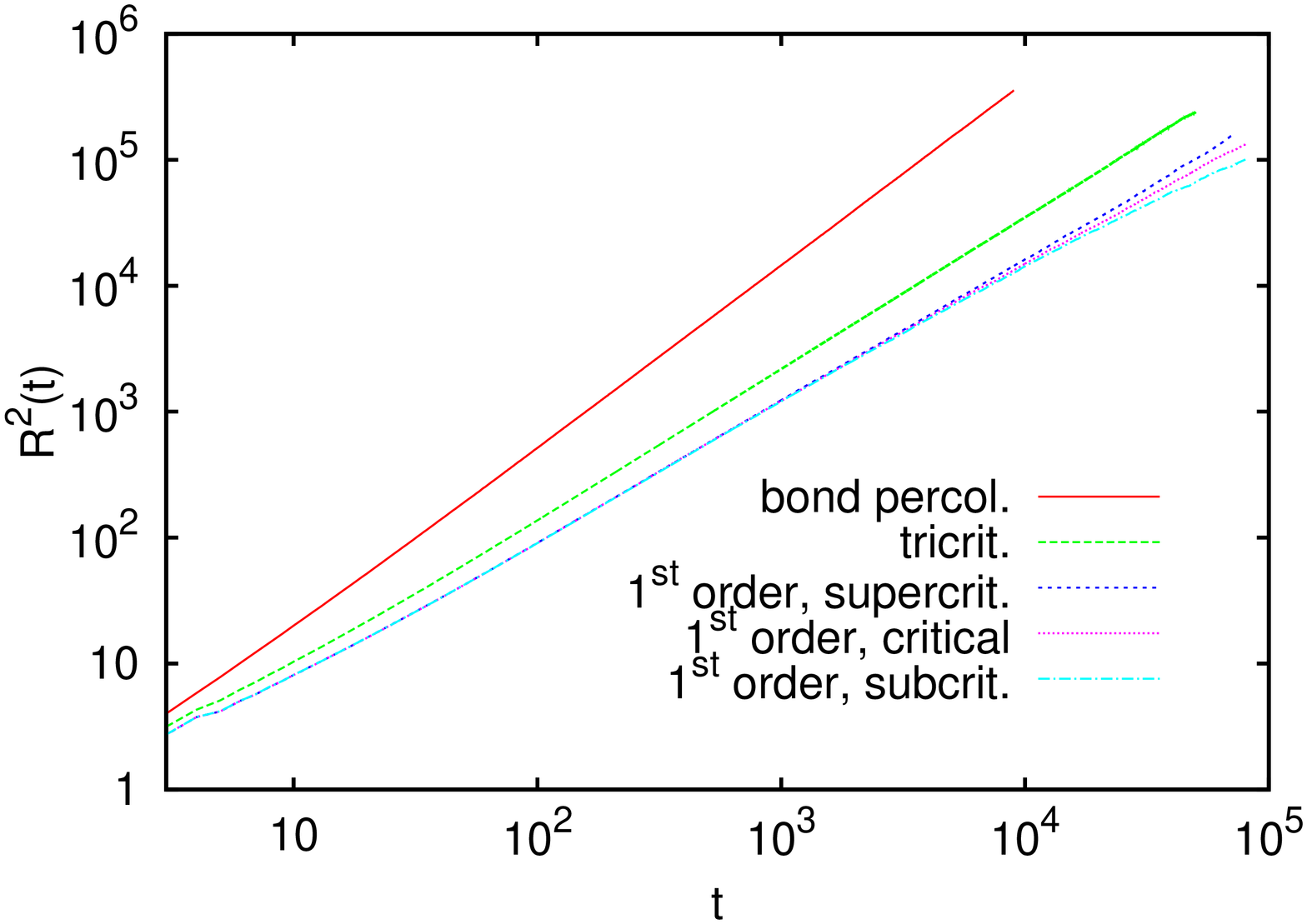,width=5.8cm, angle=270}
\caption{(Color online) Analogous to Figs.~1 and S1, but for $R^2(t)$ which is the average 
   distance of active (i.e. newly infected) sites from the seed. In the first order 
   regime the cluster grows very slowly, and $R^2(t)$ should be dominated by the motion of their center 
   of mass. Since clusters can grow only into new (not previously infected) areas, this motion 
   should be essentially a self-avoiding random walk. Our estimate for the critical exponent is 
   consistent with this, although there are large finite time corrections that make $R^2(t)$ grow
   even less fast than $t$ for very long times, in particular for $p$ below the transition. At
   the tricritical point, $R^2(t)$ is a power law with exponent $1.205(4)$.
   }
   \label{fig.S2}
   \vskip -.5 cm
\end{figure}

Further details will be given elsewhere~\cite{grass-unpub}. They will concern statistics, 
the precise methods used to estimate exact tricritical properties, critical exponents for 
rough pinned surfaces, and the behavior in dimensions different from 3. 
Here we present just two more figures (Figs.~\ref{fig.S1} and 
\ref{fig.S2}), similar to Fig.~1 of the main text, that show the survival probability $P(t)$ and the average 
squared radius $R^2(t)$ of newly infected sites, measured from the starting point 
of the epidemic. They show pwer law behavior at the tricritical point (straight 
lines on log-log plots), with exponent values given in the main text. 

A detailed comparison of our results with the field theoretic predictions of \cite{Janssen-2004} will also be given in 
\cite{grass-unpub}. Here we just mention that our results for the critical exponents are qualitatively similar 
(the changes relative to the exponents for OP go in the right directions), but the agreement if far from perfect. 
The biggest disagreement is for $\eta_s$, defined via $N(t)\sim t^{\eta_s}$. While a positive value, 
\be
   \eta_s = (\frac{1}{3}+\frac{4-\sqrt{3}}{\pi})\frac{\epsilon}{45} + O(\epsilon^2)
\ee
with $\epsilon = 5-d=2$, was predicted in \cite{Janssen-2004}, we found $\eta_s = -0.702(10)$.

\section{The epidemic threshold in the Dodds-Watts social contagion model}

Since the original derivation of the result $q_2=2q_1$ by Dodds and Watts \cite{Dodds-2004} is somewhat 
cumbersome and involves more than a minimal set of assumptions, we give here a simpler derivation following 
the typical arguments for epidemic thresholds via consistency conditions \cite{New-2001,New-2005,Son-2011}. 
We start by recalling the condition for the threshold of the standard epidemic process that leads
to ordinary percolation, and we then modify a suitable reformulation so that it also applies to the 
more general case with 
different infection probabilities $p_k$. We finally obtain the critical line and the tricritical point
by straightforward algebra.

Let us call $S$ the probability that a node at one end of a randomly chosen link gets infected during an epidemic
process on a sparse random network with degree distribution $P_k$. If the infection can pass through any link with 
probability $p$, then the locally treelike structure of the network results in the consistency condition \cite{New-2001,New-2005} 
\be
   1-S = z^{-1} \sum_{k=1}^\infty k P_k (1-p S)^{k-1}\;,     
\ee
where $z = \langle k\rangle = \sum_k k P_k$. We write this as 
\be 
   F_{\rm OP}(S) \equiv \sum_{k=1}^\infty k P_k \{(1-p S)^{k-1} + (S-1)\} = 0\   \label{S1}
\ee
(the subscripts stand for `ordinary percolation'). The percolation threshold is then defined by 
\be
   F_{\rm OP}(0) = F'_{\rm OP}(0) = 0\;,
\ee
where $F'_{\rm OP}(S) \equiv dF_{\rm OP}/dS$. Straightforward calculations give 
\cite{Boll-1985,New-2001,New-2005} 
\be
   p_c = \frac{\langle k\rangle}{\langle k(k-1)\rangle}.              \label{S2}
\ee
We also notice that $F''_{\rm OP}(0) > 0$.

In order to modify this to arbitrary infection probabilities $p_n$ for attacks following $n$ previous attacks,
we first rewrite Eq.~(\ref{S1}) such that contributions from different numbers of infected neighbors 
are separated. In order to do this we write $1-pS = (1-S) + (1-p)S$, such that the first term is the 
probability that the considered node is not infected, while the second terms is the probability that it is 
infected, but it cannot infect its neighbor since the link cannot be passed. Similarly we write
\be
   (1-p S)^k = \sum_{n=0}^{k-1} {k-1 \choose n} [(1-p)S]^n (1-S)^{k-n-1}\;,
\ee
such that each term in the sum corresponds to exactly $n$ infected neighbors. The modification to the 
generalized process is now obvious: We just have to replace the power $(1-p)^n$ by $(1-p_0)(1-p_1)\ldots
(1-p_{n-1})$. Alternatively, we can replace it by $1-q_n$ where $q_n$ is the probability that 
$n$ attacks succeed in infecting the site. The formulations using $p_n$ and $q_n$ are fully 
equivalent.

Making this modification in Eq.~(\ref{S1}) results in 
\bea 
   F_{\rm GEP}(S) & = & \sum_{k=1}^\infty k P_k \sum_{n=0}^{k-1}{k-1 \choose n} \times \\
            && \{[1-q_n]S^n (1-S)^{k-n-1} +(S-1)\} \;,  \label{S3}                  \nonumber
\eea
where GEP stands for `generalized epidemic process'. As before, the condition for criticality is 
\be
   F_{\rm GEP}(0) = F'_{\rm GEP}(0) = 0, \quad F''_{\rm GEP}(0) >0\;,
\ee
while the tricritical point is given by 
\be
   F_{\rm GEP}(0) = F'_{\rm GEP}(0) = F''_{\rm GEP}(0) =0\;,
\ee
Evaluating the derivatives is straightforward and gives \cite{Dodds-2004}
\be
   q_2 = 2q_1\;,
\ee
and $q_1$ is given by Eq.~(\ref{S2}). Notice that the location of the tricritical point does not 
depend on any $q_n$ with $n>2$, and its existence does not put any constraints on them.

In the first order regime, the threshold condition for an epidemic is 
\be 
   F_{\rm GEP}(S) = F'_{\rm GEP}(S) = 0\;,
\ee 
which depends non-trivially both on the degree distribution and on all $p_n$ (or all $q_n$). For 
regular graphs with degree $z$ (i.e., $P_k = \delta_{k,z}$) and $p_n = p$ for all $n\geq 1$ it approaches
in the limit $z\to\infty$ the linear relation $q_2 = 1/z+p_1$. For $z=15$, numerical solution of 
Eq.~(11) gives the line plotted in Fig.~ 1 of the main text.

\section{The Strauss Model}

We start from Eqs.~(5) and (6) of Ref.~\cite{Park-2005}, which read in our notation 
\be
   p = \frac{1}{e^{\theta-Bq(1-2/N)}+1}
\ee
and
\be
   q = \frac{1+(e^{B/N}-1)p}{(e^{\theta-Bq(1-3/N)}+1)^2+(e^{B/N}-1)p}\;.
\ee
Here $p$ is the link density and $q$ is defined in~\cite{Park-2005}. 
In the limit $N\to \infty$ and $B,\theta \ll N$ of interest to us, these simplify to 
\be
   p = \frac{1}{e^{\theta-Bq}+1}
\ee
and
\be
   q = \frac{1}{(e^{\theta-Bq}+1)^2} =p^2\;.
\ee
Combining these gives an equation for $p$ in terms of $B$ and $\theta$ which we can write
as 
\be
   F(p) \equiv e^{\theta-Bp^2}+1-\frac{1}{p} = 0\;.  \label{Eq.F}
\ee

\begin{figure}
\psfig{file=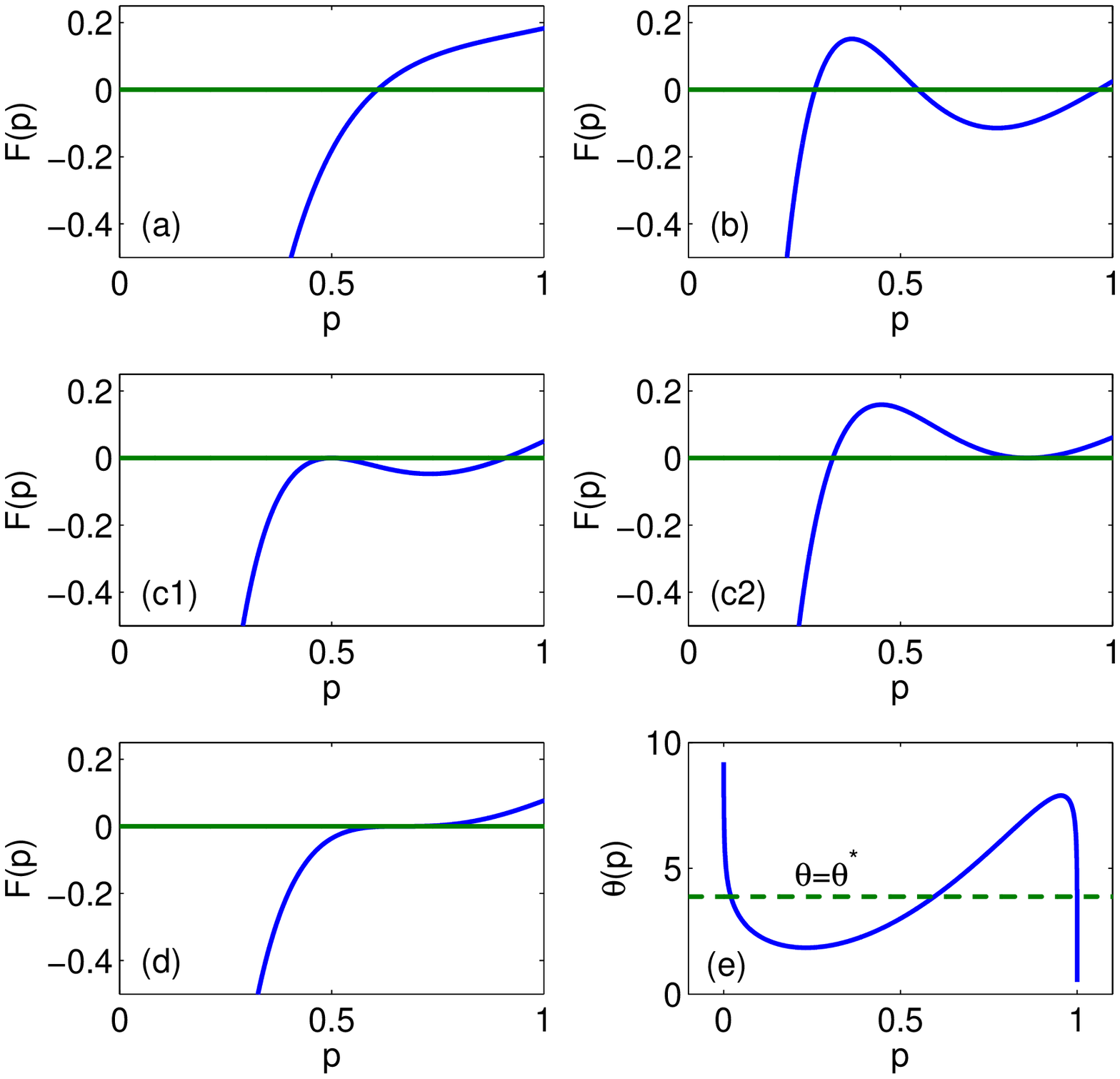,width=8.9cm, angle=0}
\caption{(Color online) Panels (a) - (d) illustrate the cases (a) - (d) for the possible solutions of 
   $F(p)=0$. Panel (e) illustrates the Maxwell construction. The function $\theta(p)$ is given by 
   Eq.~(\ref{theta}), and the dashed horizontal line is placed such that the two areas between it and 
   the curve $\theta=\theta(p)$ are equal.}
   \label{fig.strauss}
   \vskip -.5 cm
\end{figure}

This equation can have four outcomes (see Fig.~\ref{fig.strauss}):
\begin{itemize}
\item (a) one simple solution, corresponding to one single phase (i.e. no phase coexistence; Fig.~\ref{fig.strauss}a);
\item (b) three different solutions (2 stable + 1 unstable), corresponding to phase coexistence (Fig.~\ref{fig.strauss}b);
\item (c) one single plus one doubly degenerate solution, corresponding to the boundaries of the 
    coexistence region (Figs.~\ref{fig.strauss}c1 and ~\ref{fig.strauss}c2); and
\item (d) one triply degenerate solution, corresponding to the critical point (Fig.~\ref{fig.strauss}d).
\end{itemize}

The critical point is thus obtained from $F(p) = F'(p)=F''(p) = 0$, leading to the values 
quoted in the main text. The boundaries of the bistable region are given parametrically by 
\be
   B = \frac{1}{2(1-p)p^2}\;,\quad \theta = \ln (1/p-1)+\frac{1}{2(1-p)}
\ee
which give asymptotically for large $\theta$
\be
    B_-(\theta) \approx \theta\;, \quad B_+ \approx \exp(2\theta)
\ee
for the lower (upper) boundaries in a plot of $B$ versus $\theta$ (see Fig.~2 of \cite{Park-2005}).

The actual transition curve $B^*(\theta)$, or rather $\theta^*(B)$, is obtained from a Maxwell 
construction: For some given value of $B > B^*_c$, we first obtain $\theta$ as a function of $p$ from 
Eq.~(\ref{Eq.F}), 
\be
   \theta(p) = Bp^2 +\ln(1/p-1)\;.     \label{theta}
\ee
For any $\theta_0$ in the coexistence region, the equation $\theta(p) = \theta_0$ has three
roots $p_1<p_2 <p_3$. The transition point $\theta^*(B)$ is then given by 
\be
   \int_{p_1}^{p_3} dp \;\left[\theta(p) - \theta^*(B)\right] =0.
\ee
For large $B$, the curve of $\theta(p)$ versus $p$ tends (except near the points $p=0$ and $p=1$) to
a parabola, which gives in this limit $\theta^*(B)\approx B/3$ as stated in the main text.

\bibliographystyle{apsrev4-1}
\bibliography{mm}